\documentclass{JHEP3}
\usepackage{graphicx}
\usepackage{amssymb}
\usepackage[mathscr]{eucal}
\usepackage{amsmath}
\usepackage{cite}
\usepackage{afterpage}
\usepackage{slashed}
\usepackage{feynmp}

\newcommand{\gsimm}{\raise.3ex\hbox{$>$\kern-.75em\lower1ex\hbox{$\sim$}}}
\newcommand{\lsimm}{\raise.3ex\hbox{$<$\kern-.75em\lower1ex\hbox{$\sim$}}}

\newcommand{\be}{\begin{equation}}
\newcommand{\ee}{\end{equation}}
\newcommand{\ba}{\begin{eqnarray}}
\newcommand{\ea}{\end{eqnarray}}

\newcommand{\bea}{\begin{eqnarray*}}
\newcommand{\eea}{\end{eqnarray*}}

\title{Constraining Disformally Coupled Scalar Fields}
\author{Philippe Brax \\
  Institut de Physique Th\'eorique, CEA, IPhT, CNRS, URA 2306,
  F-91191 Gif/Yvette Cedex, France \\ E-mail:
  \email{philippe.brax@cea.fr}}
 \author{Clare Burrage\\
 School of Physics and Astronomy, University of Nottingham, Nottingham, NG7 2RD, United Kingdom
  \\ E-mail:
  \email{Clare.Burrage@nottingham.ac.uk} }

\date{today}
\abstract{Light scalar fields can naturally couple disformally to matter fields.  Static, non-relativistic sources do not generate a classical field profile for a disformally coupled scalar, and so such scalars are free from the constraints on the existence of fifth forces that are so restrictive for conformally coupled scalars. In this work we show that disformally coupled scalars can still be studied and constrained through their microscopic interactions with fermions and photons, both in terrestrial laboratories and from observations of stars.  The strongest constraint on the coupling scale comes from mono-photon searches at the LHC and requires $M \gtrsim 10^2 \mbox{ GeV}$.}

\begin{document}

\section{Introduction}

Are we allowed to introduce a new light scalar field \cite{Copeland:2006wr} that couples to matter \cite{Clifton:2011jh}?  For  conformally coupled scalar fields, the answer to this question appears to be ``yes, but with difficulty''.  If a scalar field has a canonical  kinetic term, and its potential consists only of a mass term, then experimental searches for fifth forces severely constrain the existence of a minimal coupling to matter\cite{Adelberger:2003zx}.  These constraints can be alleviated through screening mechanisms that introduce non-linear, higher order operators (that can be radiatively stable) but the cost is a more baroque scalar sector.  Disformal couplings present an interesting alternative; such a coupling does not (classically) result in a force between static, non-relativistic objects and therefore is not constrained by the fifth force experiments that are so restrictive for conformal couplings.

Disformal interactions were first discussed by Bekenstein \cite{Bekenstein:1992pj}, who showed that the most general metric that can be constructed from $g_{\mu\nu}$ and a scalar field that respects causality and the weak equivalence principle is;
\begin{equation}
\tilde{g}_{\mu\nu}=A(\phi,X)g_{\mu\nu} + B(\phi,X) \partial_\mu \phi \partial_\nu \phi\; ,
\label{eq:bekmetric}
\end{equation}
where the first term gives rise to conformal couplings between the scalar field and matter, and the second term is the disformal coupling.  Here  $X=(1/2)g^{\mu\nu}\partial_{\mu}\phi\partial_{\nu}\phi$.
The disformal interactions  give rise to Lagrangian interaction terms of the form
\begin{equation}
\mathcal{L} \supset \frac{1}{M^4}\partial_\mu \phi\partial_\nu \phi T^{\mu\nu}\;.
\label{eq:coupling}
\end{equation}
where $T^{\mu\nu}$ is the energy momentum tensor of matter fields. As is clear from Equation (\ref{eq:coupling}) a disformal coupling occurs through a high mass dimension operator, however if the scalar field possesses a shift symmetry then this is the lowest order operator we can write down which respects Lorentz invariance.
This direct coupling between derivatives of $\phi$ and the energy-momentum of matter is such that the matter density $T^{00}$ only couples to time derivatives of $\phi$. As a result the scalar field is not sourced by a static pressureless perfect fluid.
However, quantum effects are still present and a new force mediated by the scalar field appears at the one loop level \cite{Kugo:1999mf,Kaloper:2003yf};  we will examine this force in detail in Section \ref{sec:oneloop}.  Operators of even higher order than those in Equation (\ref{eq:coupling}) can be generated by quantum corrections.  Those involving additional derivatives of $\partial_{\mu} \phi$ give rise to terms in the equation of motion that have more than two derivatives per field and thus are expected to give rise to ghost degrees of freedom.  Such terms are suppressed by the cut off of the effective field theory which we expect to lie above the scale $M$.  Other higher order operators contain the same $\partial\phi \partial \phi T$ combination, and remain small compared to the term in Equation (\ref{eq:coupling}) if $(\partial \phi) <M^2$.

Various order of magnitude bounds on the strength of the disformal coupling were discussed in \cite{Kaloper:2003yf} the strongest of which was $M\gtrsim 200 \mbox{ GeV}$ from requiring the theory to give a unitary description of electron positron collisions at the LEP collider.  In Section \ref{sec:collider} we will show that the cross section for two fermions annihilating into two scalars takes a different form to that considered in \cite{Kaloper:2003yf}, and that therefore unitarity at LEP requires a much weaker bound on $M$. 
 Our aim in this work is to determine the best current constraints on the scale $M$.
A variety of observational probes of disformal couplings have also been previously considered: The disformal interactions of scalars with photons  can be probed in laboratory experiments \cite{Brax:2012ie}. In
models motivated by Galileon theories and massive gravity,  constraints have been
put on the disformal interactions from studying gravitational lensing and the velocity dispersion of
galaxies \cite{Wyman:2011mp,Sjors:2011iv}. A disformally coupled Galileon has been shown to fit current cosmological observations and a non-zero disformal coupling seems to be  marginally preferred \cite{Neveu:2014vua}. Other cosmological implications of disformal scalars have been considered in \cite{Zumalacarregui:2010wj,Koivisto:2012za,Bettoni:2012xv,Brax:2013nsa,vandeBruck:2013yxa}.  Disformal interactions have been shown to arise in the four dimensional effective theory resulting from various brane world scenarios \cite{deRham:2010eu,Koivisto:2013fta}, 
in branon models \cite{Alcaraz:2002iu,Cembranos:2004jp} 
 and in theories of massive gravity \cite{deRham:2010ik,deRham:2010kj}.

In this work we determine the constraints imposed on disformal scalars by considering the microscopic interactions between scalars, fermions and photons.  We will find that constraints on the disformal coupling can be imposed by a wide ranging array of laboratory experiments and astrophysical observations.
In the first section, we introduce the disformal coupling and show the absence of any classical effect in the presence of dense and non-relativistic matter.  In Section \ref{sec:collider} we consider the constraints imposed on the theory by requiring unitary evolution in particle colliders, and then bound the theory with the null results of mono-lepton searches for new physics at the LHC.   Then in Section \ref{sec:oneloop}, we investigate the quantum effects and rederive  the force between two fermions due to a scalar loop. This force is strongest at short distances therefore in Section \ref{sec:casimir} we study the macroscopic effects of the disformal interaction and consider the disformal Casimir-Polder interaction between one fermion and a plate, and the disformal Casimir effect.  In Section \ref{sec:atoms} the one loop force is  applied to atomic transitions in hydrogen-like atoms where the disformal interaction changes the atomic energy levels. We also calculate the cross section in the scattering between non-relativistic neutrons and rare gases in Section \ref{sec:neutrons}.  We then check in Section \ref{sec:stars} that the disformal interaction does not lead to a fatal increase in the burning rate of stellar structures. We conclude and summarize the constraints in Section \ref{sec:conc}.

\section{Disformally Coupled Scalar Fields}
\subsection{Effective Action}
As discussed in the introduction a disformal coupling between matter and a scalar field, $\phi$, arises because  in the Einstein frame matter fields move on geodesics of a metric $\tilde{g}_{\mu\nu}$ that depends on the scalar field.  We consider a disformal scalar field defined by the following action:
\be
S=\int d^4x \sqrt{-g}\left(\frac{R}{2\kappa_4^2} -\frac{1}{2} (\partial \phi)^2\right )  + S_m(\psi_i, \tilde g_{\mu\nu})\;,
\label{eq:action}
\ee
where the  metric is
\be
\tilde g_{\mu\nu}=  g_{\mu\nu} +\frac{2}{M^4} \partial_\mu\phi \partial_\nu \phi\;.
\label{eq:tildemetric}
\ee
This is not the most general scalar metric as given by Bekenstein in Equation (\ref{eq:bekmetric}), however it describes all the leading order effects of   disformal couplings, and is much simpler to work with.  The coupling scale $M$ is constant and unknown and should be fixed by observations.

The metric $\tilde{g}_{\mu\nu}$ is the Jordan frame metric with respect to which matter is conserved \cite{Brax:2013nsa} (this follows from diffeomorphism invariance of $S_m$):
 \be
 \tilde D_\mu \tilde T^{\mu\nu}=0\;,
\ee
where the Jordan frame energy momentum tensor is
\be
\tilde T^{\mu\nu}= \frac{2}{\sqrt{-\tilde g}} \frac{\delta S_m}{\delta \tilde g_{\mu\nu}}\;.
\ee
The Einstein frame energy-momentum for matter is $T^{\mu\nu}= (2/\sqrt{-g})(\delta S_m/\delta g_{\mu\nu})$.

\subsection{Absence of Tree Level Interactions with  Static Non-Relativistic Sources}

The disformal coupling induces interactions between the scalar field and matter at all orders in $1/M^4$.
The first order interaction reads
\be
S^{(1)}=\frac{1}{M^4} \int d^4x \sqrt{-g} \partial_\mu\phi\partial_\nu\phi T^{\mu\nu}\;.
\ee
Higher order terms are simply obtained by iteration
\be
S^{(n)}= \frac{1}{M^{4n}} \int d^4 x \sqrt{-g} C_{(n)}^{\alpha_1\beta_1 \dots \alpha_n \beta_n} (\partial_{\alpha_1}\phi\partial_{\beta_1}\phi)\dots(\partial_{\alpha_n}\phi\partial_{\beta_n}\phi)\;,
\ee
where we have identified the tensor
\be
C_{(n)}^{\alpha_1\beta_1 \dots \alpha_n \beta_n}= \left.\frac{2^n}{\sqrt{-g}}\frac{\delta^n S_m(\psi_i, \tilde g_{\mu\nu})}{\delta g_{\alpha_1\beta_1}\dots \delta g_{\alpha_n\beta_n}}\right\vert_{\tilde g_{\mu\nu}=g_{\mu\nu}}\;,
\ee
or equivalently
\be
C_{(n)}^{\alpha_1\beta_1 \dots \alpha_n \beta_n}= \left.\frac{2^{n-1}}{\sqrt{-g}}\frac{\partial^{n-1} (\sqrt{-g}T^{\alpha_1 \beta_1})}{\partial g_{\alpha_2\beta_2}\dots \partial g_{\alpha_n\beta_n}}\right\vert_{\tilde g_{\mu\nu}=g_{\mu\nu}}.
\ee
For non-relativistic matter, we have
\be
T^{\mu\nu}=\rho u^\mu u^\nu\;,
\ee
 where $\rho$ is the matter density (which is a delta function for matter particles) and the velocity 4-vector is $ u^{\mu}=dx^\mu/d\tau$
where the proper time is $d\tau= (-g_{\mu\nu}dx^\mu dx^\nu)^{1/2}$.
At second order we find that
\be
C_{(2)}^{\alpha_1\beta_1  \alpha_2 \beta_2}= (g^{\alpha_1\beta_1} + 2 u^{\alpha_1}u^{\beta_1}) T^{\alpha_2\beta_2}\;,
\ee
and by iteration, we find that all the higher order tensors are proportional to $T^{\mu\nu}$.
As a result, in static situations where $u^0=1, u^i=0$ and $\dot \phi=0$ we find that all the disformal terms vanish $S^{(n)}=0$.
This implies that the scalar field does not mediate any classical interaction between matter particles. For instance, one does not expect any interaction between test particles in a static laboratory experiment. This is not the case at the quantum level, and this is what we turn to now.

\subsection{Interactions with Fermions}
In what follows we will restrict ourselves to the leading order effects of the disformal coupling between the scalar field and matter, and so calculate only to leading order in $1/M^4$.    To this order  the action can be expanded as
\be
S=\int d^4x \sqrt{-g}\left(\frac{R}{2\kappa_4^2} -\frac{1}{2} (\partial \phi)^2 + \frac{1}{M^4} \partial_\mu\phi\partial_\nu\phi T^{\mu\nu}\right) + S_m(\psi_i, g_{\mu\nu})\;.\label{eq:GravityFieldAction}
\ee
where $T^{\mu\nu}$ is now the Einstein frame energy-momentum tensor for matter.

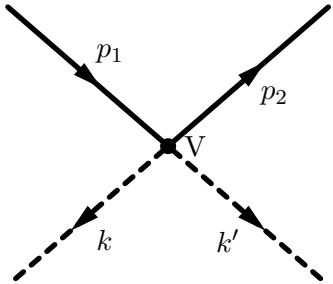
\begin{figure}
\centering
\begin{fmffile}{vertex}
\begin{fmfgraph*}(150,105)
\fmfpen{thick}
\fmfleft{i1,i2}
\fmfright{o1,o2}
\fmf{fermion,label=$p_1$}{i1,v1}
\fmf{fermion,label=$p_2$}{v1,i2}
\fmf{scalar,label=$k$}{v1,o1}
\fmf{scalar,label=$k^{\prime}$}{v1,o2}
\fmfdotn{v}{1}
\fmflabel{V}{v1}
\end{fmfgraph*}
\end{fmffile}
\caption{The disformal interaction vertex connecting two fermions (solid lines) and two scalars (dashed lines).}
\label{fig:vertex}
\end{figure}

In this paper we will focus in detail on the microscopic interactions between the scalar field and fermions, for which we first need to determine the interaction vertex.
A fermion field in the Jordan frame is characterized by the action
\be
S_F= -\int d^4 x \sqrt{-\tilde g} [i\bar \psi \tilde \gamma^\mu \tilde D_\mu \psi + m_\psi \bar\psi \psi]\;,
\ee
where the Dirac matrices and the covariant derivatives are those corresponding to $\tilde g_{\mu\nu}$.
The associated Einstein frame energy momentum tensor is given by
\be
T^{\mu\nu}_\psi= -\frac{i}{2} [\bar\psi \gamma^{(\mu}D^{\nu)}\psi - D^{(\mu} \bar\psi \gamma^{\nu)} \psi]\;,
\ee
where indices have been symmetrised and we have taken the fermions to be on-shell.
Therefore the Einstein frame scalar action contains a disformal interaction with the fermions of the form:
\be
S_{\phi}\supset-\int d^4 x \frac{i}{2M^4}\partial_\mu \phi \partial_\nu \phi [\bar\psi \gamma^{(\mu}D^{\nu)}\psi - D^{(\mu} \bar\psi \gamma^{\nu)} \psi]\;.
\ee
 In static situations where the scalar field profile is non-trivial, this implies a modification
of the fermion dispersion relation\cite{Brax:2012hm} and superluminal effects\footnote{An  analogue of Hawking's chronology protection conjecture is expected to apply to prevent the formation of closed time-like curves.  For further discussion we refer the reader to \cite{Babichev:2007dw,Burrage:2011cr}}. Here we are interested in the quantum properties of this interaction. They can be deduced from
the interaction vertex shown in Figure \ref{fig:vertex}:
\be
V=-\frac{1}{4M^4}\bar u (p_2) [(k.p_1) \slashed{k'} + (k'.p_1) \slashed{k} + (k.p_2) \slashed{k'} + (k'.p_2) \slashed{k}]u(p_1)\;,
\label{eq:vertex}
\ee
where $p_{1,2}$ are the four-momenta of the external fermions.

We now summarize our conventions for the following calculations: We take external Dirac fermions to be normalized such that
\be
\sum_s \bar u_s(p) \bar u_s(p)= -\slashed{p} +im_\psi\;,
\ee
where the sum is over the spins $\pm 1/2$. We chose  the $\gamma$ matrices to be in the Dirac representation corresponding to the mostly plus signature $(-+++)$
\be
\gamma_0=i
\left(
\begin{array}{cc}
I_2&0\\
0&-I_2\\
\end{array}\right)\;,
\ \
\gamma^i= i
\left(
\begin{array}{cc}
0&-\sigma^i\\
\sigma^i&0\\
\end{array}\right)\;,
\ee
where $I_2$ is the two dimensional identity matrix and the $\sigma^i$'s are the Pauli matrices.  This will be convenient when taking the non-relativistic limit of various interactions.

In the non-relativistic limit, we have
\be
u_s= \sqrt{2m_\psi}
\left( \begin{array} {c}
\psi_s\\
\tilde\psi_s\\
\end{array}\right )\;,
\ee
where $\tilde \psi= -\frac{\sigma.p}{m_\psi} \psi$,  the 3-momentum of the fermion is $p^i$, and $\psi_s$ is the non-relativistic wave function of a spin $1/2$ fermion.
Notice that in this limit we have $\sum_s \bar u_s(p) \bar u_s=  m_\psi\mbox{Tr}( iI_4+ \gamma^0)=2im_{\psi}$, where $I_4$ is the 4-dim identity matrix, where the last equality holds for the non-relativistic wave function of the spin $1/2$ fermions.

\section{Collider Constraints}
\label{sec:collider}
The interaction vertex shown in Figure \ref{fig:vertex} allows a fermion and an anti-fermion to annihilate into two scalars, such interactions typically occur in particle colliders, and the null results of searches for beyond the standard model physics can be used to constrain the disformal interaction.
On purely dimensional grounds the cross section for this interaction can be expected to take the form
\begin{equation}
\sigma \sim \alpha_1 \frac{E^6}{M^8} + \alpha_2 \frac{ E^4 m^2}{M^8} + \ldots
\label{eq:crosssec}
\end{equation}
where $E$ is the center of mass energy, $m$ is the mass of the fermions, and $\alpha_i$ are dimensionless coefficients.
In \cite{Kaloper:2003yf} it was assumed that the leading term in Equation (\ref{eq:crosssec}) would be the dominant contribution to the cross section, and as a result it was estimated that  requiring unitarity for electron-positron collisions at the  LEP collider imposes  $M\gtrsim 200 \mbox{ GeV}$.  Here we show that in fact $\alpha_1=0$ and therefore  the unitarity constraints on the disformal coupling scale are weaker than previously thought.

The scattering amplitude corresponding to Figure \ref{fig:vertex} is
\begin{equation}
|\mathcal{M}|^2 = \frac{1}{16 M^8}\mbox{Tr}[ \slashed{X} (-\slashed{p}_1 +im) \slashed{X} (-\slashed{p}_2 +im)]\;,
\end{equation}
where $X_{\mu} = (k\cdot p_1) k^{\prime}_{\mu}+(k^{\prime}\cdot p_1) k_{\mu}+(k\cdot p_2) k^{\prime}_{\mu}+ (k^{\prime}\cdot p_2) k_{\mu}$. We work in the center of mass frame where the incoming fermions have four-momenta $p_1= (E, \sqrt{E^2-m^2}\vec{z})$ and $p_2 = (E, - \sqrt{E^2-m^2} \vec{z})$ and the corresponding scalar momenta are $k= (E, \vec{q})$ and $k^{\prime} = (E, -\vec{q})$, where $\vec{q}^2= E^2$ and  $\vec{z}$ is a unit vector. We find that the structure of the vector $X_{\mu}$ leads to a cancellation amongst the terms that are independent of the fermion mass and as a result we find
\begin{equation}
|\mathcal{M}|^2 = \frac{8 m^2 E^6}{M^8}\;.
\label{eq:scattamp}
\end{equation}
The corresponding cross section is
\begin{equation}
\sigma = \frac{m^2 E^4}{ \pi M^8 \sqrt{1-\frac{m^2}{E^2}}  }\;.
\label{eq:discross}
\end{equation}
At energies higher than the fermion mass the square root can be expanded to put this expression into the form of Equation (\ref{eq:crosssec}), and identify $\alpha_1 =0$ and $ \alpha_2 = 1/4 \pi$.

\subsection{Unitarity}
Clearly the cross section in Equation (\ref{eq:discross}) diverges as the energy of the interaction is increased, leading to a violation of unitarity. Particle interactions have been observed to be unitary all the way up to the TeV scale energies probed by the LHC, therefore we require that $\mathcal{M}\lesssim 16 \pi$ for all observable interactions.  With the scattering amplitude of Equation (\ref{eq:scattamp}) we can update the unitarity bounds from LEP; the collider reached energies of $209\mbox{ GeV}$ when colliding electrons and positrons, for the disformal contribution to these interactions to be unitary we must impose:
\begin{equation}
M\gtrsim 3 \mbox{ GeV}\;.
\end{equation}

The LHC now reaches significantly higher energies than were accessible at LEP.  Making the conservative assumptions that the most common interactions involve up and down quarks with energies $\sim 2 \mbox{ Tev}$ then we find that unitarity requires
\begin{equation}
M \gtrsim 30 \mbox{ GeV}\;.
\end{equation}
Clearly the bounds on $M$ can be increased if heavier particles, or higher energy collisions are considered.

\subsection{Constraints from Searches for Beyond the Standard Model Physics}

The annihilation of two fermions into two scalars is difficult to detect in a particle collider.  The scalars do not decay inside the detector, and therefore we must search for a missing energy signal.  This is particularly difficult to extract from a hadron collider such as the LHC where searches rely on an observable trigger to identify an event. In searches at the ATLAS and CMS detectors fermion annihilation into undetectable particles is searched for in events where one of the incoming fermions radiates a gluon, or a jet, referred to as mono-jet searches,  or a W boson prior to the collision.  The W boson is assumed to decay into a lepton and it associated neutrino, and therefore these are known as mono-lepton searches.  The limit from mono-jet searches are very sensitive to the analysis cuts made by ATLAS and CMS, and as these cuts are done with the aim of constraining the production of WIMP dark matter particles in fermion annihilation it is unclear how to translate these bounds into constraints on disformal scalars.

Mono-lepton searches are more easily applied to disformal scalars.  The results of searches for new physics in the final states with an electron or a muon was reported by the CMS collaboration in Reference \cite{CMS:2013iea}  following the strategy of Reference \cite{Bai:2012xg}. The cross section for such a process, when the interaction is spin and flavour independent, is constrained to be
\begin{equation}
\sigma < 0.3 \mbox{ pb}\;,
\end{equation}
for light scalars.  The most common interactions involve up and down quarks, and assuming that the energy carried by these quarks can reach $2 \mbox{ TeV}$ we find that this results in a constraint on the disformal interaction of 
\begin{equation}
M\gtrsim 120 \mbox{ GeV}\;.
\end{equation}
A stronger constraint has been recently obtained using data from the ATLAS collaboration\cite{Cembranos:2013qja} and by the CMS experiment\cite{CMS} in the single photon channel with a resulting bound on the disformal coupling \footnote{We focus on the massless limit where the brane tension $f^4$  is related to the disformal coupling as $M^4=2 f^4$.}
\be
M\gtrsim 490 \mbox{ GeV}\;.
\ee

\section{The One-Loop Fifth Force}
\label{sec:oneloop}
When a source is static and non-relativistic no field profile is generated classically for a disformal scalar field at any order in $1/M$.
The absence of a force was also explicitly shown at the level of the classical equations of motion in Reference \cite{Brax:2012ie}. In References \cite{Kugo:1999mf,Kaloper:2003yf},  it was shown that the lowest order one loop diagram would correspond to a force between the fermions of the form
\begin{equation}
F\sim \frac{m_1m_2}{M^8 r^8}\;,
\end{equation}
where $m_1$ and $m_2$ are the masses of the particles being scattered.  Contributions at higher loop order are suppressed when $Mr>1$.  An estimate of the constraints from fifth force experiments  in Reference \cite{Brax:2012ie} gave $M> \mbox{MeV}$. In this section we re-derive these results, and then proceed to constrain the existence of this force with torsion balance measurements. Our derivation of the force will follow the calculation of the Casimir-Polder force presented in the textbook by Itzykson and Zuber \cite{Itzykson:1980rh}, and we will quote the main results that are derived there.

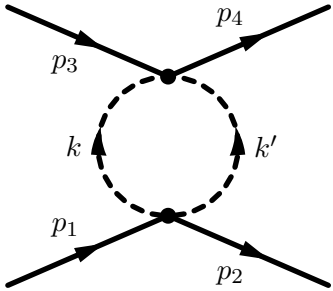
\begin{figure}
\centering
\begin{fmffile}{oneloop}
\begin{fmfgraph*}(150,105)
\fmfpen{thick}
\fmfleft{i1,i2}
\fmfright{o1,o2}
\fmf{fermion,label=$p_1$}{i1,v1}
\fmf{fermion,label=$p_2$}{v1,i2}
\fmf{fermion,label=$p_3$}{o1,v2}
\fmf{fermion,label=$p_4$}{v2,o2}
\fmf{scalar,left,tension=0.5,label=$k$}{v1,v2}
\fmf{scalar,right,tension=0.5,label=$k^{\prime}$}{v1,v2}
\fmfdotn{v}{2}
\end{fmfgraph*}
\end{fmffile}
\caption{Two fermion scattering mediated by a loop of the disformal scalar field.}
\label{fig:oneloop}
\end{figure}

We consider the four fermion interaction mediated by a loop of  two scalar fields, as shown in  Figure \ref{fig:oneloop}. At large distances and in the non-relativistic limit, the Feynman amplitude becomes a function of the momentum transfer
\be
q=p_1-p_2\;,
\ee
where momentum conservation implies that $p_1+p_3=p_2+p_4$.
In this limit we can define the interaction potential between the two fermions as
\be
V_\phi(r)=\int \slashed{d}^3 q V(q) e^{iq.r}\;,
\ee
where we use the standard notation $\slashed{d}^3 q=\frac{d^3 q}{(2\pi)^3}$.  This is related to the forward scattering amplitude ${\cal F}(q^2)$ as
\be
V(q)= i \frac{{\cal F}(q^2)}{4 m_1 m_2}\;,
\ee
where the masses of the two interacting fermions are $m_{1,2}$.

A convenient way of extracting the small $q$ (and therefore long range) behaviour of $V(q)$ is to notice that, for a fixed Mandelstam variable $s= - (p_1+p_3)^2\approx (m_1+m_2)^2$ in the non-relativistic limit, the forward amplitude $\cal F$  is an analytic function of $-q^2$ with a cut along the positive axis\footnote{ We use the $(-+++)$ signature for the metric implying that our $q^2$ corresponds to $-q^2$ in Itzykson and Zuber.}. A dispersion relation then relates $\cal{F}_R$, the real part of ${\cal F}$, on the positive axis to the forward amplitude
\be
{\cal F}(q^2)= \frac{1}{i\pi}\int_0^{\infty} dm^2 \frac{{\cal F}_R (-m^2)}{m^2+q^2 -i\epsilon}\;,
\ee
where the Cutkosky rules give
\be
{\cal F}_R(q^2)= \frac{1}{2} \int \frac{\slashed{d}^3 k}{2k^0}\frac{\slashed{d}^3 k'}{2k'^0}\slashed{\delta}^4(k+k'-q) d(k,k')\vert_{k^2=k'^2=0}\;.
\ee
Transforming back to real space we find that the non-relativistic scattering potential is then
\be
V(r)= -\frac{1}{16\pi^2 m_1 m_2r} \int_0^\infty dm^2 {\cal F}_R(-m^2)e^{-mr}\;,
\ee
where the leading order is given by the behaviour of ${\cal F}_R(q^2)$ around the origin and we have $q^2= (\vec{q})^2$.

To perform this calculation for the disformal force it is convenient to define the average
\be
\langle X\rangle= \frac{ \int \frac{\slashed{d}^3 k}{2k^0}\frac{\slashed{d}^3 k'}{2k'^0} \slashed {\delta}^4 (q-k-k') X(k,k')}{\int \frac{\slashed{d}^3 k}{2k^0}\frac{\slashed{d}^3 k'}{2k'^0} \slashed {\delta}^4 (q-k-k')}\;,
\label{eq:average}
\ee
where the denominator is a step function $\frac{1}{8\pi} \theta (-q^2)$. As a result we have that
\be
{\cal F}_R(q^2)= \frac{1}{16\pi} \theta(-q^2) \langle d\rangle\;.
\ee

The scattering amplitude corresponding to Figure \ref{fig:oneloop} reads
\be
{\cal F}(q^2)= \int {\slashed d}^4 k \slashed{d}^4 k' \frac{\slashed{\delta}^4( k+k'-q)}{(k^2 -i\epsilon)(k'^2 -i\epsilon)} d(k,k')\;,
\ee
where $\slashed {\delta}^4(k+k'-q)= (2\pi)^4\delta (k+k'-q)$ and $d$ encodes the effects of the two interaction vertices in the Feynman diagram:
\begin{align}
d(k,k')=\frac{1}{16M^8} & \bar u (p_2) [ (k.p_1) \slashed{k'} + (k'.p_1) \slashed{k} + (k.p_2) \slashed{k'} + (k'.p_2) \slashed{k}]u(p_1) \nonumber \\
 &\qquad {}\times \bar u (p_3) [ (k.p_4) \slashed{k'} + (k'.p_4) \slashed{k} + (k.p_3) \slashed{k'} + (k'.p_3) \slashed{k}]u(p_3)\;,
\end{align}
the average of which is
\begin{align}
\langle d \rangle=  \frac{(q^2)^2}{30\times 16\times M^8} & [ 2( p_1.p_3+ p_2.p_4 + p_1.p_4 + p_2.p_3) \bar u(p_2) u(p_1) \bar u(p_3) u(p_4) \nonumber \\
& \qquad {} + \bar u(p_2)( \slashed{p_3}+\slashed{p_4} )u(p_1)\bar u(p_3)( \slashed{p_1}+\slashed{p_2}) u(p_4)]\;,
\end{align}
where we have used the result that, to  leading order in $q^2$,
\be
\langle k^\mu k^\nu k'^\rho k'^\sigma \rangle= \frac{(q^2)^2}{15\times 16}(g_{\mu\nu}g_{\rho\sigma}+ g_{\mu\rho}g_{\nu\sigma}+ g_{\nu\rho}g_{\mu\sigma})\;.
\ee

In the non-relativistic limit
\be
\bar u(p_2) u(p_1) \bar u(p_3) u(p_4)=- 4 m_1m_2\;,
\ee
and
\be
\bar u(p_2)( \slashed{p_3}+\slashed{p_4} )u(p_1)\bar u(p_3)( \slashed{p_1}+\slashed{p_2}) u(p_4)= 16m_1^2m_2^2\;.
\ee
Combining all of the above  we find  that, in the non-relativistic limit,
\be
{\cal F}_R(q^2)= \frac{1}{10\pi} \theta(-q^2) \frac{m_1^2 m_2^2(q^2)^2}{16 M^8}\;.
\ee
From which we deduce that the scalar induced interaction between two fermions is
\be
V(r)= -\frac{3}{32\pi^3 r^7}\frac{m_1m_2}{ M^8}\;.
\label{eq:scalarpot}
\ee
This is an attractive interaction which is proportional to the particle masses.

We stress that although we have performed a one-loop calculation, this is the leading order potential mediated by a disformally coupled scalar between static non-relativistic objects, because for such objects the tree-level interaction vanishes at all orders in $1/M^4$.

\subsection{Constraints from Fifth Force Experiments}
\label{sec:torsion}

Constraints on deviations from a $1/r$ Newtonian gravitational potential are most commonly formulated in terms of  Yukawa corrections
 and laboratory experiments aiming to probe such corrections have been performed over a wide range of distance scales from  centimeters to tens of nanometers, for a review see \cite{Adelberger}.    Experiments probing millimeter distance scales give the tightest constraint on the magnitude of the correction to Newtonian gravity, and the constraints weaken dramatically over shorter distance scales.

The best constraints from torsion balance experiments \cite{Adelberger} find that the inverse square law holds down to a length scale of $ 56 \;\mu\mbox{m}$.  This requires that the ratio of the disformal potential to the Newtonian potential be less than unity at this distance scale.  Such a constraint requires:
\begin{equation}
0.07 \mbox{ MeV} < M\;.
\end{equation}
This is a weak bound which will be superseded by other laboratory and astrophysical constraints.

\section{The Scalar Casimir and Casimir-Polder Effects}
\label{sec:casimir}
As the disformal one-loop force scales as $\sim 1/M^8 r^8$ we still expect the tightest constraints on $M$ to come from experiments performed at the shortest possible distance scales.  These short distance experiments are for instance measurements of the Casimir force. The Casimir force is usually discussed as the pressure exerted on the bounding surfaces of a region due to the zero-point fluctuations of quantized fields in the interior.  An alternative formulation due to Jaffe \cite{Jaffe:2005vp} describes the Casimir force as the (relativistic, retarded) Van der Waals force exerted between the boundaries of a region.  In this formulation, the standard result for the Casimir force is recovered in the $\alpha \rightarrow \infty$ limit (a limit that can be shown to be appropriate on the short distance scales used to measure the Casimir effect). The Casimir-Polder effect, closely related to the Casimir force, is the force exerted on a test particle due to a nearby surface.   The Casimir force per unit area for two idealized, perfectly conducting plates with vacuum in between then scales inversely with the fourth power of the distance between the plates $F\sim 1/a^4$.  Therefore experiments searching for the Casimir force focus on probing physics at extremely short distance scales, making them ideal experiments to constrain the existence of a disformal force.

The presence of the plates does not impede the propagation of fluctuations of the scalar field, and so there is no analogue of the quantum Casimir effect for disformal scalars.  However the one-loop force calculated in Section \ref{sec:oneloop} creates an attractive force that can be constrained by searches for the Casimir and Casimir-Polder force.  These experiments are typically performed either by studying the force between two parallel plates or between a plate and a sphere.  The first case is easiest to calculate but difficult to realize experimentally.  Therefore the current best measurements of the Casimir effect come from experiments using the sphere-plate geometry.

\subsection{The Disformal Force in a Plate-Sphere Casimir Experiment}
\label{sec:casps}
The most sensitive measurement of the Casimir force are those that probe the interactions between a plate and a sphere.  To compute the disformal effects in such experiments we have to integrate the interaction  over the volume of the sphere and plate.
 We take the plate to be of density $\rho_1$ and width $a$ with infinite extent in the $(x,y)$ directions.  The shortest distance between the plate and the sphere is $d$ and the sphere has density $\rho_2$ and radius $R$.  All experiments are performed with $d \ll R$.  Points on the surface of the plate are described by  $\vec{r}_1 = (r_1\cos\theta_1, r_1\sin\theta_1,0)$ and points within the sphere have $\vec{r}_2 = (r_2\cos\theta_2\sin\phi, r_2\sin\theta_2\sin\phi,d+R-r_2\cos\phi)$, with $0\leq r_2\leq R$.  The disformal potential is then found to be
\begin{equation}
\Phi= -\frac{3\rho_1 \rho_2 a}{32 \pi^3 M^8}\int_0^{\infty}dr_1\int_0^{R}dr_2 \int_0^{2 \pi}d\theta_1\int_0^{2 \pi}d\theta_2\int_0^{\pi}d\phi \frac{r_1 r_2^2}{|\vec{r}_1-\vec{r}_2|^{7/2}}\;,
\end{equation}
whilst the full integral is difficult to compute it is clear that it is dominated by the contribution of the points of closest approach between the sphere and the cylinder. With this assumption, and  $d\ll R$, we can approximate the disformal potential as
\begin{equation}
\Phi = - \frac{3\rho_1 \rho_2a R^5}{64 \pi M^8 d^7}\;.
\end{equation}

The Casimir force between a sphere and a flat surface is
\begin{equation}
F_C = \frac{ \pi^3 R }{360 d^3}\;,
\end{equation}
and therefore the ratio of the disformal force, $F_C = \partial \Phi/\partial d$,  to Casimir force is
\begin{equation}
\frac{F}{F_C}=\frac{945}{8\pi^4}\frac{\rho_1\rho_2 a R^4}{M^8 d^5}\;.
\end{equation}

 The best constraints on the existence of a disformal force from a Casimir type experiment comes from  the measurement performed by Lamoreaux \cite{Lamoreaux:1996wh}, where $a=0.5 \mbox{ cm}$, $R=11.3 \mbox{ cm}$ and $\rho_1= \rho_2 = 2.6 \mbox{ gcm}^{-3}$.  No deviation from the theoretical prediction of the Casimir force is seen at the 5\% level when the plate and sphere are separated by a distances down to $d=0.5 \mbox{ $\mu$m}$.  This requires
\begin{equation}
0.1 \mbox{ GeV}< M\;.
\end{equation}

\subsection{The  Disformal Force in a Casimir Polder Experiment}
  In Section \ref{sec:oneloop} we proved that   two point sources are attracted by a disformal potential that scales as the inverse of the seventh power of the distance between the sources.  To compute the effect of this disformal interaction in a Casimir-Polder experiment we integrate over a uniform plate.  We approximate the experimental environment by assuming that the plate has infinite extent in the $(x,y)$ directions, and that a test particle lies a distance $z$ from the surface.  We assume that the plate has thickness $a$ and density $\rho$.
The disformal potential experienced by the test particle due to the plate is:
\begin{equation}
\Phi = - \frac{3}{32 \pi^3}\frac{1}{M^8}\int\frac{\rho}{r^7}\;dV\;,
\end{equation}
where $R^2= x^2+y^2 +z^2=r^2 +z^2$ and the integral is performed over the plate.  Therefore
\begin{eqnarray}
\Phi&=& -\frac{3}{32\pi^3}\frac{\rho a}{M^8}\int_0^{2\pi}\int_0^{\infty}\frac{r\;drd\theta}{(z^2+r^2)^{7/2}}\\
&=& -\frac{3\rho a}{80\pi^2 M^8 z^5}\;.
\end{eqnarray}

Neutrons in empty space over a thin mirror have quantized energy levels in the terrestrial gravitational field.
The disformal Casimir-Polder interaction between the neutron and the mirror changes the energy levels and the interaction potential
\be
V(z)= m \left(gz-\frac{3\rho a}{80\pi^2 M^8 z^5}\right)\;,
\ee
where $m$ is the neutron mass.
The second term must be a perturbation to the gravitational interaction as the first four energy levels of the unperturbed system have been observed to a precision of $10^{-14}$ eV \cite{Jenke:2014yel}.
The neutron energy levels in the absence of a disformal coupling are determined by the zeros of the wave functions $\psi_n (z)= c_n \mbox{Ai} \left(\frac{z}{z_0}-\epsilon_n\right)$, where Ai is the Airy function,  $z_0= \left(\frac{1}{2m_{}^2 g}\right)^{1/3}$ and $\mbox{Ai}(-\epsilon_n)=0$, resulting in the energy levels \cite{Brax:2011hb}
\be
E_n= m  gz_0 \epsilon_n\;.
\ee
The disformal coupling implies that the Casimir-Polder interaction diverges as $z\to 0$, which is not physical and corresponds to extending the validity of the effective interaction between the neutron
and the mirror to a regime where the plate cannot be considered as a dense object anymore. Indeed, the continuous plate approximation that we have used is valid only where $z\gtrsim z_{\rm atom}$ where $z_{\rm atom }\sim 10^{-10}$ m. Below this scale, the interaction becomes an interaction between individual particles and not a continuum. The approximation is valid all the way down to the atomic scale provided we have
\be
M\gtrsim \left(\frac{3\sigma}{80\pi^2 g z_{\rm atom}^6}\right)^{1/8}\approx 0.1\ \mbox{GeV}\;,
\ee
where $\sigma=\rho a\sim 17 \ \rm{gcm^{-2}}$ is the surface density of the mirror. This bound is consistent with  that previously
obtained from  Casimir experiments.

The shift in the energy levels induced by the disformal interaction becomes
\be
\delta E_n= \int_{z_{\rm atom}}^\infty dx \vert \psi_n (x)\vert^2 \frac{3m\sigma}{80\pi^2 M^8 z^5}\;.
\ee
Using $c_n^2= \frac{1}{A_n z_0}$ where $A_n= \int_{-\epsilon_n}^\infty \mbox{Ai}^2(x) dx$ and $a_n=\frac{d\mbox{Ai}}{dx}\vert_{x=-\epsilon_n}$ we find that
\be
\delta E_n \sim -\frac{3a_n^2 m\sigma}{160\pi^2 A_n z_{\rm atom}^2 z_0^3 M^8}\;,
\ee
which must be less that $10^{-14}$ eV for $n=1\dots 4$. This results in the  bound
\be
M\gtrsim 0.8 \ {\rm MeV}\;,
\ee
which is weaker than the requirement for validity of our approximations and therefore no constraint on disformal scalars can currently be applied.

\section{Constraints from Precision Atomic Measurements}
\label{sec:atoms}

Precision atomic measurements are not commonly considered tests of modifications of gravity.  However because the disformal force derived above varies as $1/r^8$, precision measurements over short distance scales can be extremely constraining, and the Bohr radius describing the size of an atom is $a_0 = (\hbar /m_e c)/\alpha= 5.3 \times 10^{-11}\mbox{ m}$, making atomic measurements potentially a very sensitive probe. Constraints from atomic measurements were placed on chameleon theories in \cite{Brax:2010gp}, and the results  of this section are derived in a similar way.

The scalar interaction acts as a perturbation of the Coulombic interaction in hydrogen-like atoms
\be
V(r)= -\frac{ e^2}{r}  -\frac{m_1m_2}{M^8} \frac{3}{32\pi^3 r^7}\;.
\label{eq:disforce}
\ee
The second, disformal, term in this expression  is strongly divergent at the origin.
As atomic precision measurements agree well with theoretical expectations (with the exception of measurements of the proton charge radius \cite{Pohl:2013yb}), the effects of the disformal force on atomic structure must be small.  In order to ensure that no modification of the electron wave functions is required we impose that the disformal perturbation must be subdominant in Equation (\ref{eq:disforce}) down to the size of the nucleon $r_N$.  This requires
\be
M^8 \gtrsim \frac{3 m_fm_N}{128\pi^4\alpha r_N^6}\;.
\ee
where $m_f$ is the mass of the fermion in the orbitals, $m_N$ is the mass of the nucleus and $\alpha$ is the fine structure constant. For a hydrogen atom this requires $M> 0.07 \mbox{ GeV}$.
In addition we will also cut off all spatial integrals at $r_N$, and assume that any divergences as $r\rightarrow 0$ are resolved by the extended size of the nucleus and its structure.

To first order  in perturbation theory, the atomic levels are perturbed by
\be
\delta E=-\frac{3m_fm_N}{32\pi^3 M^8} \left\langle E\left\vert \frac{1}{ r^7}\right\vert E\right\rangle\;,
\ee
 where $\vert E\rangle$ is the unperturbed wave function of the energy level.
 Let us focus on hydrogen-like atoms and consider the 1s, 2s and 2p levels. In each case the disformal perturbation to the  energy levels is most sensitive to the small $r$ parts of the wave function, when $r\ll a_0$ we have:
 \begin{eqnarray}
 && \psi_{1s}(r)\approx \frac{1}{\sqrt{\pi}} \left(\frac{Z}{a_0}\right)^{3/2}\;,\nonumber \\
 &&\psi_{2s}(r) \approx \frac{1}{2\sqrt{2\pi}} \left(\frac{Z}{a_0}\right)^{3/2}\;,\nonumber \\
 &&\psi_{2p}(r) \approx \frac{1}{\sqrt{\pi}} \left(\frac{Z}{2a_0}\right)^{5/2} r \cos \theta\;,\nonumber
 \end{eqnarray}
 where the Bohr radius is $ a_0= \frac{1}{m_f \alpha}$ and $Z$ is the atomic number of the nucleus.

The disformal interaction produces the following shifts in the energy levels:
 \be
 \delta E_{1s}= -\frac{3}{2^5 \pi^3} \left(\frac{Z}{a_0}\right)^{3} \frac{m_N m_f}{M^8 r_N^4}\;,
 \ee
  \be
  \delta E_{2s}= -\frac{3}{2^8\pi^3} \left(\frac{Z}{a_0}\right)^{3} \frac{m_N m_f}{M^8 r_N^4}\;,
 \ee
 \be
 \delta E_{2p}= -\frac{1}{2^9\times \pi^3} \left(\frac{Z}{a_0}\right)^{5} \frac{m_N m_f}{M^8 r_N^2}\;.
 \ee
 This leads to a disformal contribution to the lowest atomic transition $\delta E_{1s-2s}$ :
\be
 \delta E_{1s-2s}= \frac{21}{2^8\pi^3} \left(\frac{Z}{a_0}\right)^{3} \frac{m_N m_f}{M^8 r_N^4}\;.
 \ee
Similarly the
Lamb shift $\delta E_{2s-2p}= \delta E_{2p}- \delta E_{2s}$ is modified
 \be
 \delta E_{2s-2p}=\frac{3}{2^8\pi^3} \left(\frac{Z}{a_0}\right)^{3} \frac{m_N m_f}{M^8 r_N^4}\left[1- \frac{1}{6} \left(\frac{Z}{a_0}\right)^{2} r_N^2\right]\;,
 \ee
 where the second contribution in the bracket is negligible.

The most precisely measured atomic transition is the lowest, $1s-2s$, transition in hydrogen \cite{Schwob:1999zz}.  The measurement accuracy constrains   $\delta E_{1s-2s}\lesssim 10^{-9}$ eV with $Z=1$ and $m_f=m_e$ \cite{Jaeckel:2010xx}. The choice of what distance scale to take for $r_N$ is more subtle. Measurements of the Lamb shift can be interpreted as a measurement of the proton charge radius, and therefore the nuclear size for a hydrogen atom.  There is currently a significant discrepancy between measurements of this radius performed with electronic hydrogen, and those performed with muonic hydrogen \cite{Pohl:2013yb}. In this work we take the current CODATA value \cite{Mohr:2012tt} that does not include the muonic hydrogen measurements  $r_P = 0.88 \times 10^{-15} \mbox{ m}$. We discuss the charge radius measurements and their implications for disformally coupled scalars separately in more detail in \cite{us}.
 The measurement of the $1s-2s$ transition in hydrogen therefore constrains
\begin{equation}
0.2 \mbox{ GeV} < M\;.
\end{equation}

\section{Neutron Scattering Experiments}
\label{sec:neutrons}

The presence of a new force, particularly one that is strong over short distance scales, can affect the way that atoms interact with one another.  Precise constraints on the presence of such a new force come from studying  the interaction between slow neutrons and a gas, in which  the scalar field could mediate a new force between the nuclei of the gas atoms and an incoming neutron. Such experiments are performed using  thermal neutrons that scatter off noble gases such as Ne, Ar, Kr and Xe \cite{Krohn:1966zz, Nesvizhevsky:2007fv}. In our analysis we will assume that the gases are dilute enough to neglect multiple scattering, so that the problem reduces to a  2-body scattering.

The total cross section for a neutron scattering off a gas has the form:
\be
\frac{d\sigma}{d\Omega}(\theta)= \vert b+ f_{\phi}(\theta)\vert^2\;,
\ee
where $b$ is the nuclear scattering length and, in the Born approximation, the scattering amplitude due to the scalar interaction  is given  by
\be
f_{\phi }(\theta)= -2 m_N \int_{R_A}^\infty  dr r^2 \frac{\sin qr}{qr} V_{\phi} (r)\;,
\ee
where $R_A$ is the nuclear radius of atoms in the gas which  have  mass $m_A$.
In this expression $q=k\sin(\theta/2)$ and $E= k^2/2m$.
Observations constrain the  asymmetry between the forward and backward scattering cross sections. For $\theta=\pi/4$ and $\theta=3\pi/4$ this can be expressed as
\be
1+\delta R=\frac{\frac{d\sigma}{d\Omega}(\pi/4)}{\frac{d\sigma}{d\Omega}(3\pi/4)}\;,
\ee
In argon with a pressure of 1.64 atm, using $b=1.909 $  fm, $R_A \sim 3 \mbox{ fm} ^2$,  the constraint is $\delta R \le 4\times 10^{-3}$ for energies around $1$ eV \cite{Krohn:1966zz}.

The corrections due to a disformal scalar are dominated by the short distance behaviour of the disformal scalar potential, Equation (\ref{eq:scalarpot}), and we find
\begin{eqnarray}
f_{\phi }(\theta)&\approx& \frac{3m_N^2 m_A}{64\pi^2 M^8}\int_{R_A}^\infty \frac{dr}{r^5}\left(1-\frac{q^2 r^2}{6}\right)\;,\\
&\approx &\frac{3m_N^2 m_A}{256\pi^2 R_A^4 M^8}\left(1-\frac{q^2 R_A^2}{3}\right)\;,
\end{eqnarray}
where the first term renormalizes the nuclear scattering length $b$.
To leading order the resulting cross section is
\be
\frac{d\sigma}{d\Omega}(\theta)=b^2 \left(1 -\frac{m_N^2 m_A}{128\pi^2 R_A^4 M^8}\frac{q^2 R_A^2}{b}\right)\;,
\ee
and the correction to the forward-backward asymmetry is
\be
\delta R=-\frac{m_N^3 m_A}{128\pi^2 R_A^4 M^8}\frac{\sqrt 2 E  R_A^2}{b}\;.
\ee
Therefore, measurements of neutrons passing through a gas of Argon constrain:
\begin{equation}
0.03 \mbox{ GeV}<M\;.
\end{equation}

\section{ Constraints from Stellar Burning}
\label{sec:stars}
The emission of scalar fields from stars carries away additional energy, this changes the rate at which the star burns and impacts on its lifetime and structure.  We expect disformal scalars to be produced in the interior of stars through the particle interactions shown in Figures \ref{fig:brem1}-\ref{fig:pion2}, and therefore observations of the life times of the Sun, supernovae and horizontal branch stars all place constraints on the strength of the disformal coupling. An order of magnitude estimate of the constraints in \cite{Kaloper:2003yf} required $M\gtrsim 30 \mbox{ GeV}$.  In this section we consider  the different production processes in turn in the following subsections, but begin with a calculation of the mean free path of a scalar in the stellar interior.

In what follows we will make a number of approximations in order to enable us to compute the energy loss rates analytically.  Numerical simulations of the interior of stars would allow more precise bounds to be placed on the disformal scalar, this work is currently in progress \cite{withJeremy}.

\subsection{The Scalar Mean Free Path}
\begin{figure}
\centering
\begin{fmffile}{free_path}
\begin{fmfgraph*}(150,105)
\fmfpen{thick}
\fmfleft{i1,i2}
\fmfright{o1,o2}
\fmflabel{$\phi$}{i2}
\fmflabel{$\phi$}{o2}
\fmf{dbl_plain_arrow,label=$p$}{i1,v1}
\fmf{dbl_plain_arrow,label=$p_2$}{v1,o1}
\fmf{scalar,label=$k$}{i2,v1}
\fmf{scalar,label=$k^{\prime}$}{v1,o2}
\fmfdotn{v}{1}
\fmflabel{V}{v1}
\end{fmfgraph*}
\end{fmffile}
\caption{The disformal scalar scattering off a nucleon in the stellar medium.}
\label{fig:freepath}
\end{figure}
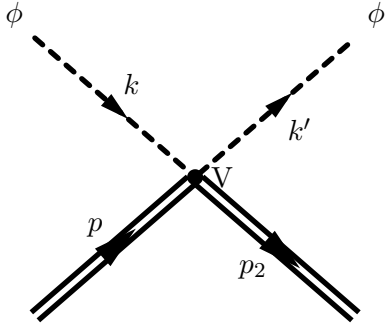

It is simplest to calculate the effects of scalar emission from stars if, once produced, the scalars escape from the star without further interaction. This happens provided the mean free path of the scalar due to the disformal interaction with fermions, shown in Figure \ref{fig:freepath},
is  larger than the size of the star.

The mean free path is related to the reaction rate between disformal scalars and nucleons in the stellar interior as
\be
\ell = \frac{1}{\Gamma}\;,
\ee
where
\begin{eqnarray}
\Gamma&=&\int \slashed{d}^3 p f_p \sigma\;,\\
\
&\sigma=&\frac{1}{2E_p 2E_k \vert \vec{v}_k-\vec{v}_p\vert} \int \frac{\slashed{d}^3 k'}{2E_k'}\frac{\slashed{d}^3 p_2}{2E_2} \slashed{\delta}^{(4)}(k'+p_2-k-p) \vert {\cal{M}}\vert^2\;,
\end{eqnarray}
and the difference of velocities is close to  unity as the scalars are massless and the fermions are non-relativistic.
The matrix element is simply
\be
{\cal M}= \bar u(p_2) V u(p)\;,
\ee
 where $V$  is the four point vertex from the disformal coupling given in Equation (\ref{eq:vertex}). In the non-relativistic approximation where $E_2\sim E_p\sim m_\psi$, the cross section is
 \be
 \sigma= \frac{m_\psi^2 E_k^4}{8\pi M^8}\;,
 \ee
  and this gives rise to a reaction  rate
\be
\Gamma= n_\psi \frac{m_\psi^2 E_k^4}{16\pi M^8}\;,
\ee
where $n_{\psi}$ is the number density of the fermion $\psi$.
Finally the mean free path is found to be:
\be
\ell= \frac{16\pi (Z+N)}{Z}\frac{M^8}{\rho T^4 m_p}\;,
\ee
where $Z$ is the atomic number of the dominant atoms in the star and $N$ is the number of neutrons.
In the Sun, where hydrogen burns predominantly we have $Z=1,\ N=0$ while  on the Horizontal Branch (HB) where Helium burns we have $Z=2,\ N=2$.
The mass of the proton is $m_p$ and we have assumed that the highest energy available to the scalars is of the order of the stellar temperature $T$.

To  estimate the mean free path for the different stars we will consider in the following, we take the most stringent lower bounds on $M$ obtained in the previous section from considerations of particle physics $M> 120 \mbox{ GeV}$.  We then find that in the Sun ($\rho\sim 150 \mbox{ g.cm}^{-3}$ and $ T\sim 1.5 \times 10^7 \mbox{ K}$)
\be
\ell_\odot \gtrsim  2\times 10^{38}\ {\rm km}\;,
\ee
where the solar radius is $R_{\odot} \sim 7 \times 10^5 \mbox{ km}$.
For stars on the horizontal branch ($\rho \sim 10^4 \mbox{ g.cm}^{-3}$ and $ T\sim 1 \times 10^8 \mbox{ K}$)
\be
\ell_{HB} \gtrsim  4\times 10^{33}\ {\rm km}\;,
\ee
where a typical radius is $R\sim 3 \times 10^7 \mbox{ km}$.
Finally for supernovae
\be
\ell_{SN}\gtrsim  10^9 \ {\rm km}\;,
\ee
larger than the progenitor solar radius.  Hence the stars are transparent to scalars interacting with matter via the disformal coupling.

\subsection{Unitarity Constraints}

All of the calculations that we  will study in this section  are perturbative and must preserve the unitarity of the underlying field theory.
In the non-relativistic approximation which is valid in all the environments from main sequence stars to supernovae, the matrix elements for the scattering a scalar off a fermion   $f\phi\to f\phi$ involving the
disformal coupling is given by
\be
{\cal M} \sim \frac{m_f^2 E^2}{M^4}
\ee
where $E$ is the energy of the incoming scalar.
Unitarity imposes that ${\cal M}\lesssim 16\pi $ where the typical energy of scalars created in the stellar medium is $E\sim T$, the temperature of the star.
We must therefore require that
\be
M\gtrsim \left(\frac{m_f T}{4 \sqrt \pi}\right)^{1/2}
\ee
for perturbative unitarity to stand. This constraint is easily satisfied in main sequence stars such as the sun where $T_\odot\sim 1.3 $ keV and $m_f=m_p$, we find
\be
M\gtrsim 31\ {\rm keV}.
\ee
For stars on the horizontal branch where $T_{HB}\sim 8.6 $ keV, we find
\be
M\gtrsim 1\ {\rm MeV}.
\ee
Finally in supernovae where $T\sim 30$ MeV we have
\be
M\gtrsim 63\ {\rm MeV}.
\ee
All these bounds are much weaker than the ones we will now derive from stellar burning rates.

\subsection{Bremsstrahlung}
\label{sec:brem}

\begin{figure}
\centering
\begin{fmffile}{brem1}
\begin{fmfgraph*}(150,105)
\fmfpen{thick}
\fmfleft{i1}
\fmflabel{$e^-$}{i1}
\fmfbottom{b1}
\fmfforce{(w/3,0)}{b1}
\fmflabel{$Ze$}{b1}
\fmfright{o1}
\fmftop{t1,t2,t3,o2,o3}
\fmflabel{$e^-$}{o1}
\fmflabel{$\phi$}{o2}
\fmflabel{$\phi$}{o3}
\fmfv{label=$V$,label.angle=-90}{v2}
\fmf{fermion,label=$p$}{i1,v1}
\fmf{fermion,label=$p_1$}{v1,v2}
\fmf{fermion,label=$p_2$}{v2,o1}
\fmf{photon,tension=0}{b1,v1}
\fmf{scalar,label=$k$,label.side=left,tension=0}{v2,o2}
\fmf{scalar,label=$k^{\prime}$,label.side=right,tension=0}{v2,o3}
\fmfdotn{v}{2}
\fmfv{decor.shape=cross}{b1}
\end{fmfgraph*}
\end{fmffile}
\caption{Bremsstrahlung of disformal scalars by an electron scattering off a nucleus.}
\label{fig:brem1}
\end{figure}

\begin{figure}
\centering
\begin{fmffile}{brem2}
\begin{fmfgraph*}(150,105)
\fmfpen{thick}
\fmfleft{i1}
\fmflabel{$e^-$}{i1}
\fmfbottom{b1}
\fmfforce{(2w/3,0)}{b1}
\fmflabel{$Ze$}{b1}
\fmfright{o1}
\fmftop{t1,t3,o2,o3,t2}
\fmflabel{$e^-$}{o1}
\fmflabel{$\phi$}{o2}
\fmflabel{$\phi$}{o3}
\fmfv{label=$\tilde{V}$,label.angle=-90}{v1}
\fmf{fermion,label=$p$}{i1,v1}
\fmf{fermion,label=$\tilde{p}_1$}{v1,v2}
\fmf{fermion,label=$p_2$}{v2,o1}
\fmf{photon,tension=0}{b1,v2}
\fmf{scalar,label=$k$,label.side=left,tension=0}{v1,o2}
\fmf{scalar,label=$k^{\prime}$,label.side=right,tension=0}{v1,o3}
\fmfdotn{v}{2}
\fmfv{decor.shape=cross}{b1}
\end{fmfgraph*}
\end{fmffile}
\caption{Bremsstrahlung of disformal scalars by an electron scattering off a nucleus.}
\label{fig:brem2}
\end{figure}

The bremsstrahlung process of scalar production is shown in Figures \ref{fig:brem1} and \ref{fig:brem2}.
We consider the emission of two scalars from the initial or final electrons interacting with the nucleus of an atom with $Z$ protons.  The matrix element corresponding to this process  can be written
as
\be
{\cal M}_T= {\cal M} +\tilde {\cal M}\;,
\ee
where ${\cal M}$ describes scalar radiation from the final state electron, and $\tilde {\cal M}$ from the initial state electron.  We have
\be
{\cal M}= \frac{Ze^2}{\vert \vec{p_1}-\vec{p}\vert^2 +m_D^2}\frac{1}{p_1^2+m_e^2} \bar u (p_2) V (i\slashed{p_1}+m_e) \gamma^0 u(p)\;,
\label{eq:curlyM}
\ee
where the disformal vertex, $V$, is given in Equation (\ref{eq:vertex}).

In the plasma where the electrons evolve, the photons are not massless  and have a mass given by the Debye mass
\be
m_D= \left(\frac{\sum_i q_i^2 n_i}{T}\right)^{1/2}=\left(\frac{4\pi \left(1+\frac{Z^2}{N+Z}\right)\alpha \rho }{m_p T}\right)^{1/2}\;,
\ee
where the charged particles are the electrons with charge $-e$ and the nuclei with charge $Ze$. The densities are $ n_e= Z n_{\rm nuclei}$ and $(N+Z) n_e=\frac{\rho}{m_p}$ where $\rho$ is the plasma density,
$m_p$ the proton mass, and $N$ the number of neutrons in the nuclei.
Numerically, we find that in the Sun we have $m_D\approx 9.5$ keV and in horizontal branch stars we have $m_D\approx 0.03 \mbox{ MeV}$. This is larger than the energy scale associated with the temperatures of the stars,  respectively $T_\odot\approx 1.3\mbox{ keV}$ and $T_{HB} \approx 8\times 10^{-3} \mbox{ MeV}$.

We define  $\tilde V$ where we exchange $p_1 \to \tilde p_1$ in Equation (\ref{eq:vertex}) and $\tilde {\cal M}$ where $p_1\to \tilde p_1$ and $V\to \tilde V$ in Equation (\ref{eq:curlyM}).  This gives
\be
\tilde {\cal M}=\frac{Ze^2}{\vert \tilde \vec{ p_1}-\vec{p_2}\vert^2 +m_D^2}\frac{1}{\tilde p_1^2+m_e^2}\bar u (p_2)\gamma^0(i\tilde{\slashed{p_1}}+m_N)\tilde V  u(p)\;.
\ee

The emission rate of disformal scalars from a star,  $\Gamma$, is given by the integral
\be
\Gamma= \frac{1}{2E v_E} \int \frac{\slashed{d}^3 k}{2E_k}\frac{\slashed{d}^3 k'}{2E_k'}\frac{\slashed{d}^3 p_2}{2E_2} \slashed{\delta}(E-E_2-E_k-E_k')(E_k+E_k')  \vert {\cal M}_T\vert^2\;,
\ee
where the velocity is $v_E=\frac{\vert \vec{p}\vert}{m_e}$, and $\vert {\cal M}_T\vert^2$ is the squared matrix element averaged over spins. The averaged energy loss rate per unit mass is given by
\be
\epsilon=\frac{n_{\rm nuclei}}{\rho} \int {\slashed{d}^3p} f_p \Gamma\;,
\ee
where $f_p$ is the thermal distribution of the initial electron which is non relativistic
\be
f_p= \frac{\rho}{2m_p} \left(\frac{2\pi}{m_e T}\right)^{3/2} e^{-\frac{\vec{p}^2}{2m_eT}}\;.
\ee
In the non-relativistic limit, the squared matrix element becomes
\be
\vert {\cal M}_T\vert^2 = \frac{64 Z^2 e^4m_e^6}{M^8} \frac{E_k^2 E_k'^2}{(p_1^2 + m_e^2)^2( \vert \vec{p_1}-\vec{p}\vert^2 +m_D^2)^2}\;.
\ee
Inserting $1= \int {\slashed{d}^4}p_1 \slashed{\delta}(p_1-p_2-k-k')$ we find that
\begin{eqnarray}
\Gamma&=& \frac{1}{2E v_E} \int \slashed{d}^3 p_1 \frac{\slashed{d}^3 p_2}{2E_2} \frac{\theta(-(p_1-p_2)^2)}{8\pi}  \langle\vert {\cal M}_T\vert^2\rangle\\
&=& \frac{1}{2E v_E} \int \slashed{d}^3 p_1 \frac{\slashed{d}^3 p_2}{2E_2} \frac{\theta(-q^2)}{10\pi}  \frac{(E-E_2) P(q^0,\vec{q})}{(p_1^2 + m_e^2)^2( \vert \vec{p_1}-\vec{p}\vert^2 +m_D^2)^2}\;,
\end{eqnarray}
where $q=p_1-p_2$ and the polynominal $P$ is given by
\be
P(q^0,\vec{q})=\frac{a}{3d} q_0^2 -\frac{2b+4c}{3d} q_0^2 q^2 + q^4\;,
\ee
where $q^2=-q_0^2 +\vec{q}^2$ and we have defined the coefficients $a=\frac{1}{30},\ b=-\frac{1}{60},\ c=\frac{1}{60}, \ d=\frac{1}{15\times 16}$ following the notation of Itzykson and Zuber \cite{Itzykson:1980rh}.
Performing the  $\vec{p_2}$, integral first we find that the  constraint $(p_1-p_2)^2\le 0$ reduces to
\be
\vert \vec{q}\vert \le  \frac{\vert \vec{p}^2 -\vec{p_1}^2\vert }{2m_e}\;,
\ee
 and therefore
\be
\Gamma= \frac{1}{2E v_E} \frac{ Z^2 e^4 m_e^5}{35 \pi^3} \int \slashed{d}^3 p_1   \frac{ \left(\frac{\vert \vec{p}^2-\vec{p_1}^2\vert}{2m_e}\right)^8}{(p_1^2 + m_e^2)^2( \vert \vec{p_1}-\vec{p}\vert^2 +m_D^2)^2}\;.
\ee
The domain of the $\vec{p_1}$ integration   is  $\vert \vec{p_1}\vert \le \frac{\vert \vec{p}^2-\vec{p_1}^2\vert}{2m_e}$ implying that $\vert \vec{p_1}\vert \le \frac{\vec{p}^2}{2m_e}$, therefore
we find that
\be
\Gamma= \frac{4 Z^2 \alpha^2 m_e}{105 \pi M^8\vert \vec{p}\vert }\left(\frac{\vec{p}^2}{2m_e}\right)^9\frac{m_e^2}{(\vec{p}^2 +m_D^2)^2}\;.
\ee
The rate per unit mass is now given by explicit integration
\be
\epsilon= \frac{ Z^2 \alpha^2m_e}{210\pi^3 A m_p} \rho \left(\frac{T}{M}\right)^8 \frac{1}{(2\pi m_e T)^{3/2}}g\left(\frac{m_D^2}{2m_e T}\right)\;,
\ee
where $A=N+Z$ and
\be
g(x)= \int_0^\infty du \frac{u^9 e^{-u}}{ \left(u+x\right )^2}\;.
\ee

For the Sun, the energy loss rate per unit mass must be $\epsilon_\odot \lesssim 0.1\ {\rm erg/s.g}$ \cite{Raffelt:1996wa}.  Taking, as before,  the temperature of the Sun to be $T\sim 1.5 \ 10^7$ K and the density $\rho\sim 150 \ {\rm g/cm^3}$, we find that this implies
\be
M \gtrsim 39 \ {\rm MeV}\;.
\ee
For stars on the horizontal branch, we have $T\sim 10^8$ K, $\rho\sim 10^4\ {\rm g/cm^3}$ and the energy loss rate due to scalars  is constrained to be $\epsilon_{HB}\lesssim 10\ {\rm erg/s.g}$. We find that this requires
\be
M_{HB}\gtrsim 173 \ {\rm MeV}\;.
\ee

\subsection{Compton Scattering}
\label{sec:compton}
\begin{figure}
\centering
\begin{fmffile}{compton}
\begin{fmfgraph*}(150,105)
\fmfpen{thick}
\fmfleft{i1}
\fmflabel{$e^-$}{i1}
\fmfright{o1}
\fmflabel{$e^-$}{o1}
\fmftop{t1,t2,t3}
\fmfforce{(w/6,h)}{t1}
\fmfforce{(5w/6,h)}{t2}
\fmfforce{(w,3h/4)}{t3}
\fmflabel{$\gamma$}{t1}
\fmflabel{$\phi$}{t2}
\fmflabel{$\phi$}{t3}
\fmfv{label=$\tilde{V}$,label.angle=-90}{v2}
\fmf{fermion,label=$p$}{i1,v1}
\fmf{fermion,label=$p_1$}{v1,v2}
\fmf{fermion,label=$p_2$}{v2,o1}
\fmf{photon,tension=0}{t1,v1}
\fmf{scalar,label=$k$,tension=0}{v2,t2}
\fmf{scalar,label=$k^{\prime}$,label.side=right,tension=0}{v2,t3}
\fmfdotn{v}{2}
\end{fmfgraph*}
\end{fmffile}
\caption{Compton process for production of disformal scalars.}
\label{fig:compton}
\end{figure}
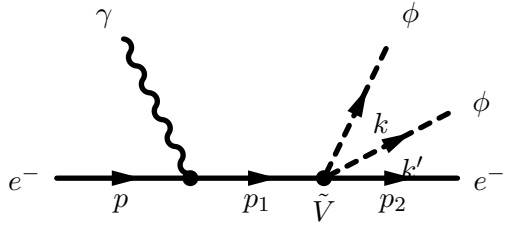

A pair of scalars can also be produced by Compton scattering between one fermion and one photon, see Figure \ref{fig:compton}.
The emission  rate is given by
\be
\Gamma= \frac{1}{2E_p 2 E_K \vert \vec {v}_E-\vec{v}_K\vert} \int \frac{\slashed{d}^3 k}{2E_k}\frac{\slashed{d}^3 k'}{2E_k'}\frac{\slashed{d}^3 p_2}{2E_2} \slashed{\delta}^{(4)}(k+k'+p_2-K-p)(E_k+E_k')  \vert {\cal M}_T\vert^2\;,
\ee
where again the velocity difference is $\vert \vec {v}_E-\vec{v}_K\vert\sim 1$ as the photons are relativistic and the fermions non-relativistic, and $\vert {\cal M}_T\vert^2$ is the squared matrix element averaged over spins and polarizations. The averaged energy loss rate per unit mass is given by
\be
\epsilon=\frac{1}{\rho} \int {\slashed{d}^3p}{\slashed{d}^3K} f_p f_K \Gamma\;,
\ee
where $f_p$ is the thermal distribution of the initial fermion which is non relativistic, and $f_K$ the photon distribution function.
The matrix element is given by
\be
{\cal M}= \frac{e}{p_1^2+m_\psi^2} \bar u(p_2) V (i\slashed{p_1} +m_\psi) \gamma^\mu \epsilon_\mu u(p)\;,
\ee
where $\epsilon^\mu$ is the photon polarization vector normalized such that
\be
\sum \epsilon_\mu\epsilon_\nu= \eta_{\mu\nu}\;,
\ee
and the sum is over the two polarizations. We have that
 \be
\vert {\cal M}_T\vert^2= -\frac{e^2}{2(p_1^2 +m_\psi^2)^2}{\rm Tr} ((\slashed{p_2} -im_\psi) V(i\slashed{p_1} +m_\psi)(\slashed{p} +2im_\psi) (i\slashed{p_1}+m_e)V)\;.
\ee
The initial and final fermions are non-relativistic while the photon spectrum is peaked around a temperature $T$ that is less than the fermion mass.
In this approximation we find that
\be
\vert {\cal M}_T\vert^2= \frac{2 e^2 m_\psi^6}{(p_1^2 +m_\psi^2)^2}\frac{E_k^2 E_k'^2}{M^8}\;,
\ee
and the emission rate  becomes
\be
\Gamma=  \int {\slashed{d}^3 q} \theta(-q^2) \frac{e^2 m_\psi^4}{32\pi(p_1^2 +m_\psi^2)^2}\frac{\langle E_k^2 E_k'^2\rangle}{M^8}\;,
\ee
where $q=p_1-p_2$. We have used the approximation $E_k+E_k'\sim E_K$. The condition $q^2\le 0$ implies that
\be
(q^0)^2\ge \vec{q}^2 = (\vec{p_1}-\vec{p_2})^2\;,
\ee
where $p_1= p+K= p_2 +k+k'$.
Using $\langle E_k^2 E_k'^2\rangle= aq_0^4 -(2b+4c)q_0^2 \vec{q}^2 +3d(\vec{q})^2$\;,
and  $X=\vert \vec{q}\vert$, we have
\be
\int {\slashed{d}^3 q} \theta(-q^2)\langle E_k^2 E_k'^2\rangle= \frac{24 d q_0^7}{7}\;,
\ee
where $d=1/15\times 16$ and  with $q_0\sim E_K$.
Now using
\be
p_1^2 +m_\psi^2 \sim -2 m_\psi E_K\;,
\ee
we find that
\be
\Gamma= \frac{3d}{28(2\pi)^3} \frac{e^2 m_\psi^2 E_K^5}{M^8}\;.
\ee
The emission rate is maximal for protons, hence we deduce the energy loss  rate due to scalars produced by Compton scattering to be
\be
\epsilon= \frac{3d h}{8\pi (2\pi)^3} \frac{\alpha Z m_p}{(Z+N)}\left(\frac{T}{M}\right)^8\;,
\ee
where
\be
h=\int_0^\infty \frac{u^7 du}{e^u-1}\;.
\ee
We find that the emission rate is smaller than that allowed  for stars on the horizontal branch when
\be
M\gtrsim 811 \ {\rm MeV}\;,
\ee
and for the Sun when
\be
M\gtrsim 236 \ {\rm MeV}\;.
\ee

\subsection{Primakov Process}
\label{sec:primakov}

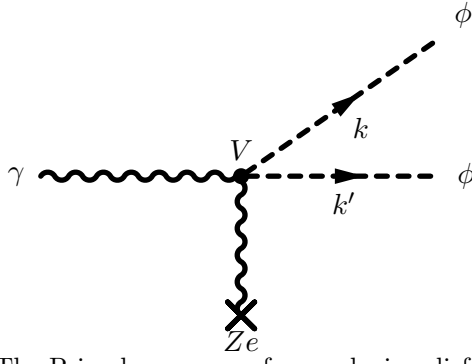
\begin{figure}
\centering
\begin{fmffile}{primakov}
\begin{fmfgraph*}(150,105)
\fmfpen{thick}
\fmfleft{i1}
\fmflabel{$\gamma$}{i1}
\fmfbottom{b1}
\fmfforce{(w/2,0)}{b1}
\fmflabel{$Ze$}{b1}
\fmfright{o1}
\fmftop{t1}
\fmfforce{(w,h)}{t1}
\fmflabel{$\phi$}{o1}
\fmflabel{$\phi$}{t1}
\fmfv{label=$V$,label.angle=90}{v1}
\fmf{photon}{i1,v1}
\fmf{photon,tension=0}{b1,v1}
\fmf{scalar,label=$k$,tension=0}{v1,t1}
\fmf{scalar,label=$k^{\prime}$}{v1,o1}
\fmfdotn{v}{1}
\fmfv{decor.shape=cross}{b1}
\end{fmfgraph*}
\end{fmffile}
\caption{The Primakov process for producing disformal scalars.}
\label{fig:primakov}
\end{figure}

We now  consider the coupling between photons and the scalar field due to the disformal term
\be
\frac{1}{M^4} \partial_\mu\phi\partial_\nu\phi \left(F^{\nu a} F^\nu_a-\frac{\eta^{\mu\nu}}{4} F^2\right)\;,
\label{eq:photonint}
\ee
which can lead to productions of scalars in the interior of a star due to the Primakov process shown in Figure \ref{fig:primakov}.
The coupling in Equation (\ref{eq:photonint}) leads to a four particle interaction vertex in the Feynman diagram expansion of perturbation theory:
\begin{align}
V_\gamma= \frac{1}{M^4} & \left( (p.k) (p'.k')(\epsilon.\epsilon')-(p.k)(k'.\epsilon')(p'.\epsilon)-(p.\epsilon')(k.\epsilon)(p'.\epsilon')\right.\nonumber \\
& \qquad {} \left.+ (p.p')(k.\epsilon)(k'.\epsilon')-\frac{(k.k')}{2}[(\epsilon.\epsilon')(p.p')-(p'.\epsilon)(p.\epsilon')]\right)\;,
\end{align}
where $\epsilon$ and $\epsilon'$ are the polarization vectors of the incoming photons of momenta $(p,p')$ and the scalars have momenta $(k,k')$.

The Primakov effect occurs when an incoming photon interacts with the electric field of a nucleus:
\be
\frac{Ze}{\vec{p'}^2 + m_D^2} \epsilon_\mu'\;,
\ee
where $\epsilon'_\mu= (1,0,0,0)$.
To simplify the calculations, we take the real and external photons to have transverse polarizations where $\epsilon_0=0$. As a result the vertex becomes
\be
V_\gamma=\frac{1}{M^4}\left[ E'_k \left((p.p')(k.\epsilon) -(p.k)(p'.\epsilon)\right)+ E_p\left(\frac{(k.k')}{2} (p'.\epsilon) -(k.\epsilon)(p'.k')\right )\right]\;.
\ee
The matrix element is simply
\be
{\cal M}= \frac{Ze}{\vec{p'}^2 + m_D^2} V_\gamma\;.
\ee
The emission rate is given by
\be
\Gamma=\frac{1}{2E_p } \int\frac{\slashed{d}^3 k}{2E_k}\frac{\slashed{d}^3 k'}{2E_k'} \slashed{\delta}^{}(E_k+E_k'-E_p) (E_k+E_k')  \vert {\cal{M}}_T\vert^2\;,
\ee
where $\vert {\cal{M}}_T\vert^2$ is the square of the matrix element averaged over the polarizations of the incoming photon.
The energy loss rate is given by
\be
\epsilon_\gamma= \frac{n_{\rm nuclei}}{\rho}\int \slashed{d}^3 p f_p \Gamma\;.
\ee
The calculation can be simplified by introducing the four-vector $p'=(0,\vec{p'})$ and inserting $1= \int \slashed{d}^3 p' \slashed{\delta}^{(3)} (\vec{p'}+\vec{p}-\vec{k}-\vec{k'})$ in $\Gamma$ so that it becomes
\be
\Gamma= \frac{1}{32\pi} \int \slashed{d}^3 p' \theta (-q^2) \frac{Z^2 e^2}{(\vec{p'}^2 + m_D^2)^2}\langle {\rm Tr}(V^2_\gamma)\rangle\;,
\ee
where the expectation value $\langle.\rangle$ was defined in Equation (\ref{eq:average}), and $q=p+p'$.
We find that
\begin{align}
{\rm Tr}(V^2_\gamma)= \frac{1}{M^8} & \left[E'^2_k \left(p'^2 (p.k)^2 -2 (p.k)(p.p')k.p'\right) +E^2_p\left(\frac{(k.k')^2}{4} p'^2 - (k.k')(p'.k')(p'.k)\right)\right.\nonumber \\
& \qquad {} \left.+2E_pE'_k\left( \frac{k.k'}{2}((p.p')(k.p')-p'^2(p.k)) +(p.k)(p'.k')(p'.k)\right)\right]\;.
\end{align}
The domain of integration is defined by $q^2\le 0$, or equivalently $\vec{p'}^2 +2 \vec{p}.\vec{p'}\le 0$. Defining the angle $\theta=(\vec{p},\vec{p'})$ between $\vec{p}$ and $\vec{p'}$, and $X=\cos\theta$, the integration can be performed over
$p'=\sqrt{\vec{p}^2}$ such that $p'\le -2 X E_p$ and then over $X$ where $-1\le X\le 0$.  In addition we approximate $\vec{p'}^2 + m_D^2\sim m_D^2$ which is justified as long as $T\lesssim m_D$ valid
for the Sun and horizontal branch stars. After an extremely lengthy calculation to obtain $\langle {\rm Tr}(V^2_\gamma)\rangle$ and the  phase space integral over $\vec{p'}$, we find that
\be
\Gamma= \frac{U Z^2 e^2}{16 (2\pi)^3 } \frac{ E_p^{11}}{m_D^4 M^8}\;,
\ee
where $U=\frac{99133}{14175}$.
The loss rate becomes
\be
\epsilon_\gamma= \frac{UhZ^2 \alpha}{4 \pi (2\pi)^3 } \frac{ T^{14}}{m_p m_D^4 M^8}\;,
\ee
where $h=\int_0^\infty dx \frac{x^{13}}{e^x-1}$.

Bounding the energy loss rate due to Primakov production of disformal scalars in  horizontal branch stars yields
\be
M\gtrsim 346 \ {\rm MeV}\;,
\ee
and  the corresponding bound from the  Sun is
\be
M\gtrsim 40 \ {\rm MeV}\;.
\ee

\subsection{Pion Exchange}
\label{sec:pion}

\begin{figure}
\centering
\begin{fmffile}{pion1}
\begin{fmfgraph*}(150,105)
\fmfpen{thick}
\fmfleft{i1,i2}
\fmfforce{(0,h/2)}{i2}
\fmfforce{(0,0)}{i1}
\fmfright{o1,o2}
\fmfforce{(w,h/2)}{o2}
\fmfforce{(w,0)}{o1}
\fmftop{t1,t2,t5,t3,t4}
\fmflabel{$\phi$}{t3}
\fmflabel{$\phi$}{t4}
\fmfv{label=$V$,label.angle=-90}{v3}
\fmf{fermion,label=$K$}{i1,v1}
\fmf{fermion,label=$p$}{i2,v2}
\fmf{scalar,label=$\pi$,tension=0}{v1,v2}
\fmf{fermion,label=$K_2$}{v1,o1}
\fmf{fermion,label=$p_1$}{v2,v3}
\fmf{fermion,label=$p_2$}{v3,o2}
\fmf{scalar,label=$k$,tension=0}{v3,t3}
\fmf{scalar,label=$k^{\prime}$,label.side=right,tension=0}{v3,t4}
\fmfforce{(w/3,0)}{v1}
\fmfdotn{v}{3}
\end{fmfgraph*}
\end{fmffile}
\caption{Production of disformal scalars by pion exchange.}
\label{fig:pion1}
\end{figure}

\begin{figure}
\centering
\begin{fmffile}{pion2}
\begin{fmfgraph*}(150,105)
\fmfpen{thick}
\fmfleft{i1,i2}
\fmfforce{(0,h/2)}{i2}
\fmfforce{(0,0)}{i1}
\fmfright{o1,o2}
\fmfforce{(w,h/2)}{o2}
\fmfforce{(w,0)}{o1}
\fmftop{t1,t2,t5,t3,t4}
\fmflabel{$\phi$}{t5}
\fmflabel{$\phi$}{t3}
\fmfv{label=$\tilde{V}$,label.angle=-90}{v3}
\fmf{fermion,label=$K$}{i1,v1}
\fmf{fermion,label=$p$}{i2,v3}
\fmf{scalar,label=$\pi$,tension=0}{v1,v2}
\fmf{fermion,label=$K_2$}{v1,o1}
\fmf{fermion,label=$\tilde{p}_1$}{v3,v2}
\fmf{fermion,label=$p_2$}{v2,o2}
\fmf{scalar,label=$k$,label.side=left,tension=0}{v3,t5}
\fmf{scalar,label=$k^{\prime}$,label.side=right,tension=0}{v3,t3}
\fmfforce{(2w/3,0)}{v1}
\fmfdotn{v}{3}
\end{fmfgraph*}
\end{fmffile}
\caption{Production of disformal scalars by pion exchange.}
\label{fig:pion2}
\end{figure}

The interiors of supernovae differ substantially from the interiors of main sequence and horizontal branch stars. The dominant process for producing disformal scalars becomes  the creation of two scalars with the exchange of one pion between two nuclei. The diagrams are depicted in Figures \ref{fig:pion1} and \ref{fig:pion2}.
As before we define the emission  rate
\be
\Gamma= \frac{1}{2E_p 2E_K  v_p} \int \frac{\slashed{d}^3 k}{2E_k}\frac{\slashed{d}^3 k'}{2E_k'}\frac{\slashed{d}^3 p_2}{2E_2}\frac{\slashed{d}^3 K_2}{2E_{K2}} \slashed{\delta}^4(p_2+K_2+k+k'-p-K)(E_k+E_k')  \vert {\cal N}_T\vert^2\;,
\ee
and the rate per unit mass
\be
\epsilon=\frac{1}{\rho} \int {\slashed{d}^3p} {\slashed{d}^3K}f_p f_K \Gamma\;.
\ee
Fortunately, in the non-relativistic limit, the rate per unit mass from the pion exchange can be related to the previous bremsstrahlung calculation.
Indeed the pion propagators  reduce to $\frac{1}{(\vec{k}+\vec{k'})^2 +m_\pi^2}$ and the nucleon propagator is simply $\frac{1}{(\vec{k}+\vec{k'})^2}$ implying that one can factorize the propagators
and write
\be
{\cal N}_T= \frac{A^2 g_{\pi NN}^2}{((p_1-p)^2 +m_\pi^2)(p_1^2+m_N^2)}[ {\cal N}_1 +{\cal N}_2+ {\cal N}_3+ {\cal N}_4]\;,
\ee
where the number of nucleons per nucleus is $A=Z+N$. We consider that the nucleons interact coherently, and we have defined
\begin{eqnarray}
{\cal N}_1&=&    (\bar u (p_2) V (i\slashed{p_1}+m_N) \gamma^0 u(p))\bar u(K_2) u(K)\;,\nonumber \\
{\cal N}_2&=&   (\bar u (p_2)  (i\tilde {\slashed{p_1}}+m_N)\tilde V \gamma^0 u(p))\bar u(K_2) u(K)\;,\nonumber \\
{\cal N}_3&=&  (\bar u (K_2) W (i \slashed{K_1}+m_N) \gamma^0 u(K))\bar u(p_2) u(p)\;,\nonumber \\
{\cal N}_4&=&    (\bar u (K_2)  (i\tilde {\slashed{K_1}}+m_N)\tilde W \gamma^0 u(K))\bar u(p_2) u(p)\;.\nonumber \\
\end{eqnarray}
The scalar-pion vertices that we label $W$ and $\tilde W$ follow the same pattern as $V$ and $\tilde V$ in Section \ref{sec:brem} with the $p$ momenta replaced by $K$'s. We find that
\be
\vert {\cal N}_T\vert^2 = \frac{2^9 A^4 g_{\pi NN}^4m_N^8}{M^8} \frac{E_k^2 E_k'^2}{(p_1^2 + m_N^2)^2( \vert \vec{p_1}-\vec{p}\vert^2 +m_\pi^2)^2}\;,
\ee
from which we can directly obtain that
\be
\Gamma= \frac{4 A^4 \alpha_\pi ^2 m_N}{210 \pi M^8\vert \vec{p}\vert }\left(\frac{\vec{p}^2}{2m_N}\right)^9\frac{m_N^2}{(\vec{p}^2 +m_\pi^2)^2}\;,
\ee
where $\alpha_\pi=\frac{g^2_{\pi NN}}{4\pi}\sim 13.5$.
Using $\int \slashed{d}^3 K f_K= \frac{n_N}{2}= \frac{\rho}{2m_p}$  the total number of nucleons divided by 2 (where we take all the nucleons to have the proton mass),  we find that
\be
\epsilon= \frac{ A^4 \alpha_\pi^2}{105\pi^3 } \rho \left(\frac{T}{M}\right)^8 \frac{1}{(2\pi m_N T)^{3/2}}g\left(\frac{m_\pi^2}{2m_N T}\right)\;.
\ee

Observations of supernova SN1987A, constrain the energy loss due to any new physics to be  $\epsilon_{SN}\lesssim 10^{19} \rm {erg/s.g}$. Taking the supernova to have a temperature of  $T=30$ MeV and a density $\rho\sim 3\times 10^{14}\ {\rm g/cm^3}$, we find that
\be
M\gtrsim 92 \ {\rm GeV}\;.
\ee
However this should be considered an order of magnitude estimate only as the nuclear forces are strongly coupled with a large coupling constant $\alpha_{\pi NN}$ and so higher order effects not computed here may become important.
Moreover, the emission rate $\epsilon$ is also only an estimate as it involves a certain number of astrophysical uncertainties.

\section{Summary and Conclusions}
\label{sec:conc}

\subsection{Summary of Constraints}

In Table \ref{tab:summary} we summarize the constraints on disformal couplings derived in this paper, and give a reference to the section in which each constraint is derived.  The most constraining observations are the null results of mono-photon searches for beyond the Standard Model physics performed by the CMS collaboration.   We present each constraint with a comment on the environment it is derived in, as in some theories with disformal couplings, such as the Galileon \cite{Nicolis:2008in}, the coupling scale can be renormalized by an environmentally dependent factor.  We hope this will allow the reader to
apply these results to their preferred theory with a disformal coupling.

\begin{table}
\begin{tabular}{ |c | c | c | c| }
  \hline
  Source of bound & Lower bound on $M$ in GeV & Environment & Discussed in Section \\
	  \hline
		Unitarity at the LHC & 30 &Lab. vac. & \ref{sec:collider}\\
		CMS mono-lepton & 120 &Lab. vac. & \ref{sec:collider}\\
   CMS mono-photon & 490 &Lab. vac. & \ref{sec:collider}\\
  Torsion Balance & $7 \times 10^{-5}$  & Lab. vac.& \ref{sec:torsion}\\
Casimir effect & 0.1 & Lab. vac.  & \ref{sec:casps}\\
  Hydrogen spectroscopy & 0.2  & Lab. vac.  & \ref{sec:atoms} \\
 Neutron scattering & 0.03  & Lab. vac.& \ref{sec:neutrons} \\
Bremsstrahlung & $4 \times 10^{-2}$ & Sun &\ref{sec:brem}\\
  & 0.18  & Horizontal Branch & \ref{sec:brem}\\
Compton Scattering & 0.24  & Sun &\ref{sec:compton}\\
 & 0.81 & Horizontal Branch & \ref{sec:compton}\\
Primakov & $4\times 10^{-2}$  & Sun &\ref{sec:primakov}\\
   & 0.35 & Horizontal Branch & \ref{sec:primakov}\\
Pion exchange & $\sim 92$  & SN1987a & \ref{sec:pion}\\
  \hline
\end{tabular}
\caption{Summary of the constraints on the disformal coupling scale $M$. Lab. vac. means the constraint derives from a laboratory vacuum on Earth.  Horizontal branch means the constraint derives from observations of horizontal branch stars, and similarly for constraints labelled Sun.}
\label{tab:summary}
\end{table}

\subsection{Conclusion}

Nearly massless scalar fields can play a role in either generating the right amount of energy to trigger the late time acceleration of the Universe or in modifying gravity on very large scales. In this paper, we have considered the possible coupling between such a scalar and matter which preserves a shift symmetry. This symmetry is instrumental in building models such as Galileons or K-mouflage \cite{Babichev:2009ee} and guarantees the absence of potential for the scalar field, i.e. that the field is massless. The classical  disformal coupling of a scalar field to matter cannot be tested in static situations  as  no tree-level force between static objects is generated.  Nevertheless the theory can still be constrained.  We have shown that collider searches for new physics give the most stringent bounds on the disformal coupling.   We have shown that  quantum mechanical effects at the one loop level  lead to a disformal force that could have consequences in atomic physics. The disformal interaction also plays a role in the heart of stellar objects where the disformal coupling opens up new channels for their burning rates. We have calculated the energy emission rates due to the disformal interaction in stars on the main sequence and the horizontal branch, and found stringent bounds on the disformal coupling strength.

The astrophysical effects following our new calculation on the disformal burning rates of stars can be applied to study the disformal effects on the Hertzsprung-Russell diagram of stellar structures; this is under study.
In another publication, we also explore in further detail the applications of the disformal coupling to atomic physics.

\section*{Acknowledgements}
We would like to thank Christoph Englert for discussions about collider  constraints on disformal scalars.  We would also like to thank Jose Cembranos and Jeremy Neveu for helpful communications.  C.B. is
supported by a Royal Society University Research Fellowship. P.B.
acknowledges partial support from the European Union FP7 ITN
INVISIBLES (Marie Curie Actions, PITN- GA-2011- 289442) and from the Agence Nationale de la Recherche under contract ANR 2010
BLANC 0413 01.

¥

\begin{thebibliography}{unsrt}


\bibitem{Copeland:2006wr}
  E.~J.~Copeland, M.~Sami and S.~Tsujikawa,
  Int.\ J.\ Mod.\ Phys.\ D {\bf 15} (2006) 1753
  [hep-th/0603057].

\bibitem{Clifton:2011jh}
  T.~Clifton, P.~G.~Ferreira, A.~Padilla and C.~Skordis,
  Phys.\ Rept.\  {\bf 513} (2012) 1
  [arXiv:1106.2476 [astro-ph.CO]].

\bibitem{Adelberger:2003zx}
  E.~G.~Adelberger, B.~R.~Heckel and A.~E.~Nelson,
  Ann.\ Rev.\ Nucl.\ Part.\ Sci.\  {\bf 53} (2003) 77
  [hep-ph/0307284].


\bibitem{Bekenstein:1992pj}
  J.~D.~Bekenstein,
  Phys.\ Rev.\ D {\bf 48} (1993) 3641
  [gr-qc/9211017].
  
\bibitem{Kugo:1999mf}
  T.~Kugo and K.~Yoshioka,
  Nucl.\ Phys.\ B {\bf 594} (2001) 301
  [hep-ph/9912496].

\bibitem{Kaloper:2003yf}
  N.~Kaloper,
  Phys.\ Lett.\ B {\bf 583} (2004) 1
  [hep-ph/0312002].
  


\bibitem{Brax:2012ie}
  P.~Brax, C.~Burrage and A.~-C.~Davis,
  JCAP {\bf 1210} (2012) 016
  [arXiv:1206.1809 [hep-th]].

\bibitem{Wyman:2011mp}
  M.~Wyman,
  Phys.\ Rev.\ Lett.\  {\bf 106} (2011) 201102
  [arXiv:1101.1295 [astro-ph.CO]].

\bibitem{Sjors:2011iv}
  S.~Sjors and E.~Mortsell,
  JHEP {\bf 1302} (2013) 080
  [arXiv:1111.5961 [gr-qc]].
	
	\bibitem{Neveu:2014vua}
  J.~Neveu, V.~Ruhlmann-Kleider, P.~Astier, M.~Besançon, A.~Conley, J.~Guy, A.~Möller and N.~Palanque-Delabrouille {\it et al.},
  arXiv:1403.0854 [gr-qc].

\bibitem{vandeBruck:2013yxa}
  C.~van de Bruck, J.~Morrice and S.~Vu,
  Phys.\ Rev.\ Lett.\  {\bf 111} (2013) 161302
  [arXiv:1303.1773 [astro-ph.CO]].

\bibitem{Brax:2013nsa}
  P.~Brax, C.~Burrage, A.~-C.~Davis and G.~Gubitosi,
  JCAP {\bf 1311} (2013) 001
  [arXiv:1306.4168 [astro-ph.CO]].

\bibitem{Zumalacarregui:2010wj}
  M.~Zumalacarregui, T.~S.~Koivisto, D.~F.~Mota and P.~Ruiz-Lapuente,
  JCAP {\bf 1005} (2010) 038
  [arXiv:1004.2684 [astro-ph.CO]].

\bibitem{Koivisto:2012za}
  T.~S.~Koivisto, D.~F.~Mota and M.~Zumalacarregui,
  Phys.\ Rev.\ Lett.\  {\bf 109} (2012) 241102
  [arXiv:1205.3167 [astro-ph.CO]].

\bibitem{Bettoni:2012xv}
  D.~Bettoni, V.~Pettorino, S.~Liberati and C.~Baccigalupi,
  JCAP {\bf 1207} (2012) 027
  [arXiv:1203.5735 [astro-ph.CO]].

\bibitem{deRham:2010eu}
  C.~de Rham and A.~J.~Tolley,
  JCAP {\bf 1005} (2010) 015
  [arXiv:1003.5917 [hep-th]].

\bibitem{Koivisto:2013fta}
  T.~Koivisto, D.~Wills and I.~Zavala,
  arXiv:1312.2597 [hep-th].
  
\bibitem{Alcaraz:2002iu}
  J.~Alcaraz, J.~A.~R.~Cembranos, A.~Dobado and A.~L.~Maroto,
  Phys.\ Rev.\ D {\bf 67} (2003) 075010
  [hep-ph/0212269].  
  
\bibitem{Cembranos:2004jp}
  J.~A.~R.~Cembranos, A.~Dobado and A.~L.~Maroto,
  Phys.\ Rev.\ D {\bf 70} (2004) 096001
  [hep-ph/0405286].   
  
  

\bibitem{deRham:2010kj}
  C.~de Rham, G.~Gabadadze and A.~J.~Tolley,
  Phys.\ Rev.\ Lett.\  {\bf 106} (2011) 231101
  [arXiv:1011.1232 [hep-th]].


  
  
  

\bibitem{deRham:2010ik}
  C.~de Rham and G.~Gabadadze,
  Phys.\ Rev.\ D {\bf 82} (2010) 044020
  [arXiv:1007.0443 [hep-th]].

\bibitem{Brax:2012hm}
  P.~Brax,
  Phys.\ Lett.\ B {\bf 712} (2012) 155
  [arXiv:1202.0740 [hep-ph]].

\bibitem{Babichev:2007dw}
  E.~Babichev, V.~Mukhanov and A.~Vikman,
  JHEP {\bf 0802} (2008) 101
  [arXiv:0708.0561 [hep-th]].

\bibitem{Burrage:2011cr}
  C.~Burrage, C.~de Rham, L.~Heisenberg and A.~J.~Tolley,
  JCAP {\bf 1207} (2012) 004
  [arXiv:1111.5549 [hep-th]].

\bibitem{CMS:2013iea}
  CMS Collaboration [CMS Collaboration],
pp collision events at center-of-mass energy of 8 TeV ,''
  CMS-PAS-EXO-13-004.
	
\bibitem{Bai:2012xg}
  Y.~Bai and T.~M.~P.~Tait,
  Phys.\ Lett.\ B {\bf 723} (2013) 384
  [arXiv:1208.4361 [hep-ph]].
  
  
\bibitem{Cembranos:2013qja}
  J.~A.~R.~Cembranos, R.~L.~Delgado and A.~Dobado,
  Phys.\ Rev.\ D {\bf 88} (2013) 075021
  [arXiv:1306.4900 [hep-ph]].
  
\bibitem{CMS}
CMS Collaboration, CMS-PAS-EXO-12-047 .


\bibitem{Itzykson:1980rh}
  C.~Itzykson and J.~B.~Zuber,
  New York, Usa: Mcgraw-hill (1980) 705 P.(International Series In Pure and Applied Physics)

\bibitem{Adelberger}
E.~G.~Adelberger, J.~H.~Gundlach, B.~R.~Heckel, S.~Hoedl, S.~Schlamminger,
 Progress in Particle and Nuclear Physics, {\bf 62}, (2009),  102-134,
[http://dx.doi.org/10.1016/j.ppnp.2008.08.002].

\bibitem{Jaffe:2005vp}
  R.~L.~Jaffe,
  Phys.\ Rev.\ D {\bf 72} (2005) 021301
  [hep-th/0503158].

\bibitem{Lamoreaux:1996wh}
  S.~K.~Lamoreaux,
  Phys.\ Rev.\ Lett.\  {\bf 78} (1997) 5
   [Erratum-ibid.\  {\bf 81} (1998) 5475].

\bibitem{Jenke:2014yel}
  T.~Jenke, G.~Cronenberg, J.~Burgdorfer, L.~A.~Chizhova, P.~Geltenbort, A.~N.~Ivanov, T.~Lauer and T.~Lins {\it et al.},
  Phys.\ Rev.\ Lett.\  {\bf 112} (2014) 151105
  [arXiv:1404.4099 [gr-qc]].

\bibitem{Brax:2011hb}
  P.~Brax and G.~Pignol,
  Phys.\ Rev.\ Lett.\  {\bf 107} (2011) 111301
  [arXiv:1105.3420 [hep-ph]].

\bibitem{Brax:2010gp}
  P.~Brax and C.~Burrage,
  Phys.\ Rev.\ D {\bf 83} (2011) 035020
  [arXiv:1010.5108 [hep-ph]].

\bibitem{Pohl:2013yb}
  R.~Pohl, R.~Gilman, G.~A.~Miller and K.~Pachucki,
  Ann.\ Rev.\ Nucl.\ Part.\ Sci.\  {\bf 63} (2013) 175
  [arXiv:1301.0905 [physics.atom-ph]].


 \bibitem{Krohn:1966zz}
  V.~E.~Krohn and G.~R.~Ringo,
  Phys.\ Rev.\  {\bf 148} (1966) 1303.

  \bibitem{Nesvizhevsky:2007fv}
  V.~V.~Nesvizhevsky, G.~Pignol and K.~V.~Protasov,
  arXiv:0705.4478 [nucl-ex].


\bibitem{Schwob:1999zz}
  C.~Schwob, L.~Jozefowski, B.~de Beauvoir, L.~Hilico, F.~Nez, L.~Julien, F.~Biraben and O.~Acef {\it et al.},
  Phys.\ Rev.\ Lett.\  {\bf 82} (1999) 4960.

\bibitem{Jaeckel:2010xx}
  J.~Jaeckel and S.~Roy,
  Phys.\ Rev.\ D {\bf 82} (2010) 125020
  [arXiv:1008.3536 [hep-ph]].

\bibitem{Mohr:2012tt}
  P.~J.~Mohr, B.~N.~Taylor and D.~B.~Newell,
  Rev.\ Mod.\ Phys.\  {\bf 84} (2012) 1527
  [arXiv:1203.5425 [physics.atom-ph]].

\bibitem{us}
P. Brax, C. Burrage,
To appear.

\bibitem{withJeremy}
P. Brax, C. Burrage, J. Sakstein,
To appear.

\bibitem{Raffelt:1996wa}
  G.~G.~Raffelt,
  Chicago, USA: Univ. Pr. (1996) 664 p


\bibitem{Nicolis:2008in}
  A.~Nicolis, R.~Rattazzi and E.~Trincherini,
  Phys.\ Rev.\ D {\bf 79} (2009) 064036
  [arXiv:0811.2197 [hep-th]].
	
\bibitem{Babichev:2009ee}
  E.~Babichev, C.~Deffayet and R.~Ziour,
  Int.\ J.\ Mod.\ Phys.\ D {\bf 18} (2009) 2147
  [arXiv:0905.2943 [hep-th]].
























	
	
	
	
\end{thebibliography}
\end{document}